\documentclass[prd, preprint, superscriptaddress, amsmath, amssymb, nofootinbib]{revtex4}
\usepackage{graphicx}
\usepackage{dcolumn}
\usepackage{bm}
\usepackage{amssymb}
\usepackage{amsmath}
\usepackage{epsfig}    
\usepackage{color}
\usepackage{slashed}
\usepackage{hhline}
\usepackage{changes}

\def\be{\begin{equation}}
\def\ee{\end{equation}}
\newcommand{\bea}{\begin{eqnarray}}
\newcommand{\eea}{\end{eqnarray}}
\newcommand{\nn}{\nonumber}

\begin{document}

\begin{flushright} APCTP Pre2022-004
\end{flushright}

\title{  A model explaining the
new CDF II W boson mass linking to  muon $g-2$ and dark matter}

\author{Keiko I.  Nagao}
\email{nagao@ous.ac.jp}
\affiliation{Okayama University of Science,  Faculty of Science,   Department of Applied Physics,  Ridaicho 1-1,  Okayama,  700-0005,  Japan}

\author{Takaaki Nomura}
\email{nomura@scu.edu.cn}
\affiliation{College of Physics,  Sichuan University,  Chengdu 610065,  China}

\author{Hiroshi Okada}
\email{hiroshi.okada@apctp.org}
\affiliation{Asia Pacific Center for Theoretical Physics,  Pohang 37673,  Republic of Korea}
\affiliation{Department of Physics,  Pohang University of Science and Technology,  Pohang 37673,  Republic of Korea}

\date{\today}

\begin{abstract}
 We propose a model to explain  the W boson mass anomaly reported by CDFII collaboration that would suggest new physics (NP).
We introduce exotic fermions; one isospin doublet vector-like lepton,  one isospin singlet singly-charged vector-like lepton, and an isospin doublet inert scalar.  
The proposed model provides sizable muon anomalous magnetic moment (muon $g-2$) due to no chiral suppression and nonzero mass difference between the real and imaginary parts of neutral inert scalar bosons. 
The inert scalar mass squared difference and vector-like exotic leptons ($L' , \ E'$ in the main text) affect oblique parameters.
Especially, $T$-parameter shift from zero
explains the W boson mass anomaly. 
We search for the allowed parameter region to explain both muon $g-2$ and W boson mass anomaly at the same time. 
We also discuss a dark matter (DM) candidate assuming the real part of the inert scalar field to be the one. 
We find that lighter DM mass is favored to be consistent with experimental constraints. 
\end{abstract}
\maketitle

\section{Introduction}
Recently,  CDFII collaboration reported an updated measurement of the standard model (SM) charged-gauge boson (W boson) mass~\cite{CDF:2022hxs}
\begin{align}
m_W = (80. 433\pm 0. 0064_{\rm stat}\pm 0. 0069_{\rm syst})\ {\rm GeV}, 
\end{align}
which deviates from the SM prediction by 7$\sigma$,  where the SM tells us $m_W = (80. 357\pm 0. 006)\ {\rm GeV}$.  
In the previous global combination of data from LEP,  CDF,  D0,  and ATLAS experiments,  the $W$ boson mass is estimated in the range of $m_W = (80. 379\pm 0. 012)\ {\rm GeV}$~\cite{ParticleDataGroup:2018ovx}. 
Here we call the discrepancy between the CDF II result and the SM prediction of W boson mass as the (CDF II) W boson mass anomaly. 
This anomaly suggests NP beyond the SM~\cite{Fan:2022dck,  Lu:2022bgw,  Athron:2022qpo,  Yuan:2022cpw,  Strumia:2022qkt,  Yang:2022gvz,  deBlas:2022hdk,  diCortona:2022zjy,  Du:2022pbp,  Tang:2022pxh,  Blennow:2022yfm,  Cacciapaglia:2022xih,  Sakurai:2022hwh,  Fan:2022yly,  Liu:2022jdq,  Lee:2022nqz,  Cheng:2022jyi,  Song:2022xts,  Bagnaschi:2022whn,  Paul:2022dds,  Bahl:2022xzi,  Asadi:2022xiy,  DiLuzio:2022xns,  Athron:2022isz,  Gu:2022htv,  Heckman:2022the,  Babu:2022pdn,  Heo:2022dey,  Du:2022brr,  Cheung:2022zsb,  Crivellin:2022fdf,  Endo:2022kiw,  Biekotter:2022abc,  Balkin:2022glu,  Krasnikov:2022xsi,  Ahn:2022xeq,  Han:2022juu,  Zheng:2022irz,  Kawamura:2022uft,  Ghoshal:2022vzo,  Perez:2022uil,  Kanemura:2022ahw,  Mondal:2022xdy,  Zhang:2022nnh,  Borah:2022obi,  Chowdhury:2022moc,  Arcadi:2022dmt,  Cirigliano:2022qdm,  Carpenter:2022oyg,  Popov:2022ldh,  Ghorbani:2022vtv,  Du:2022fqv,  Bhaskar:2022vgk,  Batra:2022org,  Cao:2022mif,  Zeng:2022lkk,  Baek:2022agi,  Borah:2022zim,  Almeida:2022lcs,  Cheng:2022aau,  Heeck:2022fvl,  Addazi:2022fbj,  Lee:2022gyf,  Cai:2022cti,  Benbrik:2022dja,  Yang:2022qgs,  Batra:2022pej,  Tan:2022bip,  Abouabid:2022lpg,  Chen:2022ocr,  Zhou:2022cql,  Gupta:2022lrt,  Basiouris:2022wei,  Wang:2022dte,  Botella:2022rte,  Barman:2022qix,  Kim:2022hvh,  Li:2022gwc,  Isaacson:2022rts,  Evans:2022dgq,  Chowdhury:2022dps,  Ghosh:2022zqs}, and
can be interpreted as the deviation of oblique parameters~\cite{Peskin:1991sw, Peskin:1990zt},  that are defined by quantum corrections to bilinear terms associated with electroweak gauge fields; especially $\Delta T$, which is a shift of oblique $T$-parameter, is important since it is related to change of relation between $W$ and $Z$ boson masses. 
In fact,  the oblique parameters are zero in the SM and new particles are required to induce non-zero values of them. 

In this letter,  we explain the W boson mass anomaly by introducing exotic fields; one isospin doublet vector-like fermion $(L')$,  one isospin singlet singly-charged vector-like fermion ($E$) and an isospin doublet inert scalar ($\eta$).  
Besides explaining the anomaly,  we can apply these new fields to obtain sizable contribution to muon anomalous magnetic dipole moment,  denoted by muon $g-2$, 
where its experimentally observed value is also deviated from the SM prediction, $a_\mu^{\rm SM}$, by 4.2$\sigma$ level.~\cite{Muong-2:2021ojo,  Aoyama:2012wk, Aoyama:2019ryr, Czarnecki:2002nt, Gnendiger:2013pva, Davier:2017zfy, Keshavarzi:2018mgv, Colangelo:2018mtw, Hoferichter:2019mqg, Davier:2019can, Keshavarzi:2019abf, Kurz:2014wya, Melnikov:2003xd, Masjuan:2017tvw, Colangelo:2017fiz, Hoferichter:2018kwz, Gerardin:2019vio, Bijnens:2019ghy, Colangelo:2019uex, Blum:2019ugy, Colangelo:2014qya, Hagiwara:2011af}. 
In particular,  the new fermions play an important role to obtain sizable contribution to muon $g-2$ since a one-loop diagram associated with new fermion is not suppressed by light lepton masses; such a diagram has no chiral suppression.  
On the other hand,  
the doublet fermion negatively contributes to $\Delta T$ while the doublet boson can positively contribute to it in this model.  
Since the W boson mass anomaly indicates the positive value of $\Delta T$, the doublet scalar is definitely required.
Moreover, the doublet scalar field includes a DM candidate which is also good motivation to introduce NP. 
Therefore, we can explain the W boson mass anomaly and inconsistency between experiments and theory for muon $g-2$ and DM simultaneously. 

This letter is organized as follows. 
In Sec.  II,  we review the model and show masses and mixings for new fields. 
In Sec.  III,  
we formulate NP contributions to the oblique parameters $\Delta T$ and $\Delta S$,  muon $g-2$,  decay width of $h\to \gamma\gamma$ mode,  relic density of DM,  and discuss direct detection constraints of DM in the model. 
In Sec. IV,  we carry out numerical analysis and demonstrate allowed parameter regions. 
Sec.  V is devoted to our summary and conclusion.

\section{Model setup and Constraints}
\begin{table}
\begin{tabular}{|c|c|c|c|c||c|c|}\hline\hline  
& ~$L_{L_i}$~  & ~$\ell_{R_i}$~& ~$L'$~& ~$E$~  & ~$H$ ~&~ $\eta$~  \\\hline\hline 
$SU(3)_C$ & $\bm{1}$  & $\bm{1}$ & $\bm{1}$ & $\bm{1}$ & $\bm{1}$ & $\bm{1}$  \\\hline 
$SU(2)_L$ & $\bm{2}$  & $\bm{1}$ & $\bm{2}$ & $\bm{1}$ & $\bm{2}$ & $\bm{2}$  \\\hline 
$U(1)_Y$   & $-\frac12$ & $-1$  & $-\frac12$& $-1$ & $\frac12$ & $\frac12$  \\\hline
$Z_2$ & $+$ & $+$ & $-$ & $-$ &  $+$ & ${-}$  \\\hline
\end{tabular}
\caption{ 
Charge assignments of the SM leptons $L_{L_i},\ \ell_{R_i}$ and the new fields  $L'\equiv [N', E']^T$,  $E$, and $\eta$ 
under $SU(3)_C\times SU(2)_L\times U(1)_Y\times Z_2$ symmetry.  
Here $L'$ and $E$ are vector-like fermions and the lower indices $i$ of $L_L,\ \ell_R$ run over $e,\mu,\tau$.
$\eta$ is an inert scalar field,  and its real neutral component is considered as a DM candidate. }
\label{tab:1}
\end{table}

In this section,  we define the model, 
introducing the following new field contents into the SM: 
\begin{itemize}
\item a vector-like exotic lepton $L'\equiv [N', E']^T$ (weak isospin doublet) ,  
\item a singly charged-lepton $E$ (weak isospin singlet), 
\item an inert  doublet scalar field $\eta$. 
\end{itemize}
The neutral component of $\eta$ can be a DM candidate. 
Furthermore,  we impose odd parity under $Z_2$ discrete symmetry for new fields that assure the stability of DM. 
%
The Higgs field in the SM
is denoted by $H$ and its vacuum expectation value (VEV) is written by $\langle H \rangle = [0, v/\sqrt2]^T$. 
The field contents and their charge assignments are summarized in Tab.~\ref{tab:1}. 
Then,  the renormalizable Lagrangian 
is given by 
\begin{align}
-\mathcal{L}_{}
=&
y_{\ell_{ij}} \overline{ L_{L_{i}}} H \ell_{R_{j}}+ y_{L'_i} \overline{ E_R}  \eta^\dagger L_{L_i} 
+ y_{S_i} \overline{\ell_{R_i}}  \eta^\dagger L'_L   + y_E \overline{L'} H E  + M_{E} \overline {E} E  +  M_{L'} \overline{ L'} L'
+ {\rm h. c. }, 
\label{Eq:lag-flavor}
\end{align}  
where $i, j=e, \mu, \tau$ are the flavor indices and  we have abbreviated kinetic terms for simplicity.
$y_\ell,\ y_E,\ M_E,\ M_{L'}$ are real while $y_{L'},\ y_S$ are complex, but we assume these couplings to be real.

The scalar potential 
is written as follows
\begin{align}
V &=  -M_H^2 H^\dagger H + M^2_\eta\eta^\dag\eta  
+ \lambda_H (H^\dagger H)^2 + \lambda_\eta(\eta^\dag\eta )^2 + \lambda_{H \eta} (H^\dagger H)(\eta^\dag\eta)\nn\\
&
 + \lambda'_{H \eta} (H^\dagger \eta)(\eta^\dag H) 
 + [\lambda''_{H \eta} 
 (H^\dagger \eta)^2+{\rm h. c. }],
\end{align}
where all parameters except $\lambda''_{H \eta} $ are real without loss of generality, but we presume it is real.
Here we explicitly write components of $H$ and $\eta$ to be
\begin{align}
H = 
\begin{pmatrix}
w^+ \\
\frac{1}{\sqrt{2}} (v + {h} + i G_Z ) 
\end{pmatrix}, \
\eta = 
\begin{pmatrix}
\eta^+ \\
\frac{1}{\sqrt{2}} ({\eta_R} + i \eta_I) 
\end{pmatrix}, 
\label{eq:scalar-fields}
\end{align}
where ${h}$  corresponds to CP-even physical scalar boson state while charged component $w^+$ and neutral
CP-odd component $G_Z$ are the Nambu-Goldstone(NG) bosons which are absorbed by the SM weak gauge bosons $W$ and $Z$,  respectively. 
The components in $\eta$ are all physical  where each of $\eta^+$ and $\eta_{R, I}$ corresponds to charged and neutral scalar bosons. 
We consider $\eta_{R, I}$ to be DM that requires $\eta$ to have zero VEV and there is no mixing between $\eta$ and $H$ to keep $Z_2$ symmetry unbroken; thus components in $\eta$ are inert scalar bosons. 
Here,  we simply define masses of  $\eta_R$,  $\eta_I$ and $\eta^\pm$ to be $m_{R}$,  $m_I$ and $\ m_{\eta^\pm}$,  respectively. 
The differences among inert scalar boson masses arise from terms with couplings $\lambda'_{H \eta}$ and $\lambda''_{H \eta}$ such that 
\begin{equation}
m_R^2 - m_I^2 = \lambda''_{H \eta} v^2,  \quad m^2_{\eta^\pm} - m^2_{R} = - (\lambda'_{H \eta} + \lambda''_{H \eta}) v^2. \label{eq:mdif}
\end{equation}
One finds that the mass squared difference between $\eta_R$ and $\eta_I$ affects the size of muon $g-2$ and the scattering cross section of DM that is tested by direct detection searches while the one between $\eta^\pm$ and $\eta_R$ induces the oblique parameter $T$ through one-loop diagrams as discussed later.  


Exotic singly-charged fermions $(E, E')$ mix each other after spontaneous electroweak symmetry breaking,  and the corresponding mass matrix is written by
\begin{align}
{\cal M}_E=
\left(\begin{array}{cc}
M_E & m_E \\
m_E &M_{L'}\\
\end{array}\right),  
\end{align}
where  $m_E\equiv y_E v /\sqrt2$. 
The mass matrix can be diagonalized by a unitary matrix $V_E$ as ${\rm diag}(M_1, M_2)=V_E^\dag {\cal M}_E V_E$, 
 and $V_E$ is given by
\begin{align}
V_E=
\left(\begin{array}{cc}
c_c & -s_c \\
s_c &c_c \\
\end{array}\right),   \quad \tan 2 \theta_c = \frac{2 m_E}{M_{E} - M_{L'}}, 
\label{eq:mixing}
\end{align}
where $c_c(s_c)$ is the shorthand notation of $\cos\theta_c(\sin\theta_c)$. 
We write mass eigenstates as $E_{1}$ and $E_2$ whose masses are denoted as $M_{1}$ and $M_2$; original states and mass eigenstates are related as $(E_1,  E_2)^T = V_E (E,  E')^T$. 

\section{Phenomenological Constraints and implications} 

In this section,  we summarize the formulas for NP contributions to
 oblique parameters,  muon $g-2$,  the partial width of the SM Higgs decay into diphoton, and
DM annihilation cross section induced by Yukawa interactions. 
We also discuss direct detection searches of DM and consider a way how to evade these constraints.  

\subsection{Oblique parameters \label{sec:oblique}}
The vector-like fermions $L'(E)$ and an inert scalar doublet $\eta$ contribute to oblique parameters via vacuum polarization diagrams for electroweak gauge bosons. 
Here,  we compute
new contributions of $S$- and $T$-parameters which are
given as follows:~\cite{Peskin:1990zt,  Ko:2021lpx} 
\begin{align}
\label{eq:S}
& \alpha_{} S^{\rm VLF} = 4 e^2 \left( \sum_{\rm VLF} \left[ \frac{d}{d q^2} \Pi_{33}^{\rm FF'} - \frac{d}{d q^2} \Pi_{3Q}^{\rm FF'} \right]_{q^2 = 0}
 \right),  \\ 
 \label{eq:T}
& \alpha_{} T^{\rm VLF} = \frac{e^2}{s_W^2 c_W^2 m_Z^2} \left( \sum_{\rm VLF} \left[ \Pi^{\rm FF'}_{\pm}(q^2) - \Pi^{\rm FF'}_{33}(q^2) \right]_{q^2 =0} 
\right), 
\end{align}
where {the} superscript VLF is shorthand notation of vector-like fermion,  $FF'$ on $\Pi^{ {\rm FF'}}_{(33, 3Q, \pm)}$  expresses the possible different combination of the vector-like fermions $F(F') = \{E_1,  E_2,  N'\}$,
%
and $\alpha \approx 1/137$ is the electromagnetic fine-structure constant. 
Explicit forms of $\Pi^{FF'}_{33, 3Q, \pm}$ can be obtained by calculating vacuum polarization diagrams and they are listed in Appendix \ref{appdx:polarizationdiagrams}. 
The contributions of $\eta$ to $S$- and $T$-parameters are simply given by~\cite{Barbieri:2006dq}
 \begin{align}
\Delta S^\eta &\approx \frac{1}{2\pi} \int_0^1 dx x(1-x) \ln\left[\frac{x m_R^2+(1-x) m_I^2}{m_{\eta^\pm}^2}\right], \\
\Delta T^\eta &\approx \frac{1}{24\pi^2 \alpha_{\rm em} v^2}(m_{\eta^\pm} - m_I)(m_{\eta^\pm} - m_R). 
\end{align}
In total,  we obtain deviation of $S$- and $T$-parameters from the SM prediction as
\begin{equation}
\Delta S = \Delta S^{\rm VLF} + \Delta S^\eta,  \quad \Delta T = \Delta T^{\rm VLF} + \Delta T^\eta. 
\end{equation}
In the numerical analysis,  we calculate these  deviations 
and compare the value with the region preferred to explain the W boson mass anomaly that is provided by ref.~\cite{Strumia:2022qkt}.

\subsection{ Muon anomalous magnetic dipole moment}
New results on the muon $g-2$ were recently published by the E989 collaboration at Fermilab \cite{Muong-2:2021ojo}: 
\begin{align}
a^{\rm FNAL}_\mu =116592040(54) \times 10^{-11}, 
\label{exp_dmu}
\end{align}
where $\mu$ indicates muon. 
Combined with the previous BNL result,  this result indicates that the muon $g-2$ deviates from the SM prediction, $a_\mu^{\rm SM}$, by 4.2$\sigma$ level~\cite{Muong-2:2021ojo,  Aoyama:2012wk, Aoyama:2019ryr, Czarnecki:2002nt, Gnendiger:2013pva, Davier:2017zfy, Keshavarzi:2018mgv, Colangelo:2018mtw, Hoferichter:2019mqg, Davier:2019can, Keshavarzi:2019abf, Kurz:2014wya, Melnikov:2003xd, Masjuan:2017tvw, Colangelo:2017fiz, Hoferichter:2018kwz, Gerardin:2019vio, Bijnens:2019ghy, Colangelo:2019uex, Blum:2019ugy, Colangelo:2014qya, Hagiwara:2011af}. 
The deviation from the SM prediction is given by
\begin{align}
\Delta a^{\rm exp}_\mu =
a^{\rm FNAL}_\mu -  a_\mu^{\rm SM}= 
 (25. 1\pm 5. 9)\times 10^{-10} , 
\label{exp_dmu}
\end{align}
where $\Delta a^{\rm exp}_\mu$ is the difference between the experimental value and the SM prediction that cannot be explained by the SM framework. 

\begin{figure}[tb]
\begin{center}
\includegraphics[width=77.0mm]{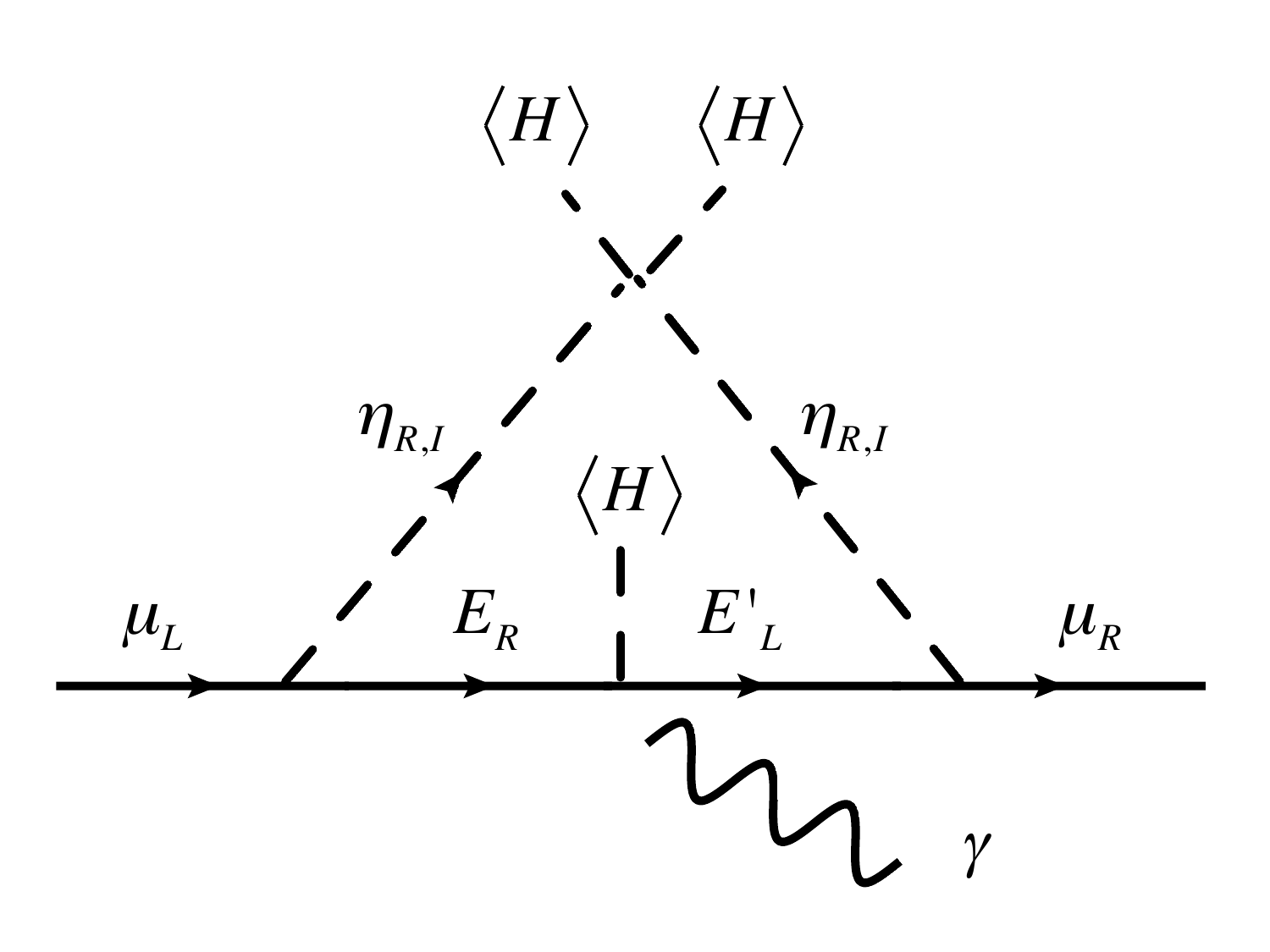} 
\caption{One loop diagram dominantly contributing to muon $g-2$. 
} 
  \label{fig:diagram}
\end{center}\end{figure}

In this model,  we expect that $\Delta a^{\rm exp}_\mu$
can be explained by NP contribution which comes from the one loop diagrams involving the vector-like leptons and neutral components of $\eta$. 
The dominant contribution is given by the diagram in Fig.~\ref{fig:diagram} and consists of  
a product of $y_{L'_\mu} $ and $y_{S_\mu}$ without chiral suppression~\cite{Crivellin:2021rbq, Crivellin:2018qmi, Calibbi:2018rzv, Sabatta:2019nfg, Benbrik:2015evd}~\footnote{The contributions proportional to $|y_{L'}|^2$ or $|y_S|^2$ cannot be dominant because of the chiral suppression. Thus, we neglect those terms for simplicity. }.  
%
NP contribution to muon $g-2$ is found as
\begin{align}
\Delta a^{\rm new}_\mu &=\frac{m_\mu}{(4\pi)^2}s_c c_c y_{L'_\mu} y_{S_\mu}
\sum_{a=1, 2}\frac{(-1)^{a-1}}{M_a}
\left[ {F(q_a,  r_{Ra})} -  {F(q_a,  r_{Ia})} \right] , \\
F(q, r)&\equiv \int_0^1 dx\int_0^{1-x} dy
\frac{1-2y}{1-x + q^2 (x^2-x) + r^2 x  }
\approx
\frac{-1+r^4-2r^2\ln(r^2)}{2(-1+r^2)^3} +{\cal O}(q^2), 
\label{exp_dmu}
\end{align}
where $q_i\equiv \frac{m_\mu}{M_i}(<<1)$ and $r_{ai} \equiv \frac{m_{a}}{M_i}$ ($a=R, \ I$; $i=1, 2$).  
As can be seen from the above formula,  the new contribution exactly vanishes when $m_R=m_I$. 
Thus,  the (not so small) mass difference between $\eta_R$ and $\eta_I$ is crucially important to obtain a sizable discrepancy from the SM muon $g-2$.
In numerical analysis,  we take the experimental region,  given in Eq.~\eqref{exp_dmu}, at 3$\sigma$ interval that is $[25.1 - 3 \times 5.9, \ 25.1 + 3 \times 5.9] \times 10^{-10}$.

\subsection{Modification of $h \to \gamma \gamma$ decay width}

New charged particles in the model modify the diphoton decay width of the SM Higgs boson. 
Relevant interactions are given by
\begin{equation}
    \mathcal{L} \supset \frac{y_E}{\sqrt{2}} h (\overline {E'_L} E_R + \overline{E}_R E'_L )+ \lambda_{H\eta} v h \eta^+ \eta^-, 
\end{equation}
where $E$ and $E'$ are written by mass eigenstates through mixing  Eq.~\eqref{eq:mixing}. 
We obtain deviation from the SM prediction via one loop diagrams in which $\{\eta^\pm,  E_1,  E_2\}$ propagate inside the loop,  such that 
\begin{align}
&\mu_{\gamma \gamma} \equiv \frac{\Gamma(h \to \gamma \gamma)}{\Gamma(h \to \gamma \gamma)_{\rm SM}} 
\simeq \left| 1 + \frac{C_E \left(  \frac{A_{1/2} (\tau_{1} )}{M_1} -\frac{A_{1/2} (\tau_{2} )}{M_2}  \right)
+ \frac{\lambda_{H \eta} v^2}{2 m_{\eta^\pm}^2 } A_0(x_{\eta^\pm})  }{A_1(\tau_W) + \frac43 A_{1/2}(\tau_t) 
 } \right|^2,   \\
& C_E =  \frac{s^2_c c^2_c (M_E - M_{L'} )}{c^2_c - s^2_c},   \  A_{1/2}(x) = -2 [x+(1-x) f(x)],  \ A_1(x) = 2+3x+3(2x-x^2)f(x),  \nonumber 
\end{align}
where $f(x) = [\sin^{-1} (1/\sqrt{x})]^2$ for $x>1$ and $\tau_X = 4m_X^2/m^2_h$ with $m_h$ being the SM Higgs boson mass. 
The current bound is given by ATLAS~\cite{ATLAS:2020qdt} and CMS~\cite{CMS:2020gsy} collaborations as
\begin{align}
\mu_{\gamma \gamma}^{\rm ATLAS} = 1. 06^{+0. 08}_{-0. 07},   \quad
\mu_{\gamma \gamma}^{\rm CMS} = 1. 01^{+0. 09}_{-0. 14},  \label{eq:hto2photons}
\end{align}
where we adopt an assumption that we do not have an exotic Higgs decay mode and only diphoton coupling is modified.

\subsection{ Dark matter}  
In this paper, we fix the DM to be $\eta_R$. It implies that negative $\lambda''_{H\eta}$ is chosen from Eq.~(\ref{eq:mdif}). Furthermore,
we expect $y_{L'_\mu} \gg y_{L'_e}, y_{L'_\tau}$ and $y_{S_\mu} \gg y_{S_e}, y_{S_\tau}$ to enhance the muon $g-2$. 
Under this assumption, thermally averaged cross section of DM annihilation to explain the current relic density is expanded in terms of the squared relative velocity of DM,  $v_{rel}^2\approx 0. 2$, 
as $ \langle\sigma v_{\rm rel}\rangle\approx a_{\rm eff}+b_{\rm eff} v^2_{\rm rel}+\cdots$, 
where $a_{\rm eff}$ and $b_{\rm eff}$ are respectively the coefficients of $s$-wave and $p$-wave,  and $\langle\cdots\rangle$ represents the thermal average. 
In this case,  the annihilation cross section is $p$-wave dominant,   
and the contribution by the Yukawa terms is given by~\cite{Chiang:2017zkh} 
\begin{align}
\langle\sigma v_{\rm rel}\rangle_Y&\simeq
\frac{|y_{L'_\mu}|^4}{392\pi} 
\left[ 
s_c^4 \frac{m_R^2}{(m_R^2+M_{1}^2)^2}
+
c_c^4 \frac{m_R^2}{(m_R^2+M_{2}^2)^2}
+
\frac{m_R^2}{(m_R^2+M_{L'}^2)^2}
\right]  v^2_{\rm rel}
\nn\\
&+
\frac{|y_{S_\mu}|^4}{392\pi} 
\left[ 
c_c^4 \frac{m_R^2}{(m_R^2+M_{1}^2)^2}
+
s_c^4 \frac{m_R^2}{(m_R^2+M_{2}^2)^2}
\right]  v^2_{\rm rel}. 
\label{eq:relic-deff2}
\end{align}
We expect the annihilation cross sections from the Yukawa terms 
to be sub-dominant.
In addition, we consider the effect of Higgs potential (except the pole at around half of Higgs mass $\sim63$ GeV) not dominant due to the constraint from the direct detection searches. 
In fact,  we find the annihilation cross section  can reach $10^{-10}$ GeV$^{-2}$ at most, which is below the value of cross section to explain observed relic density by one order magnitude through the numerical estimation.  
Therefore,  we expect the current relic density to be given via gauge interactions such as $2\eta_R\to W^+W^-/ ZZ^*$ via $\eta^\pm/\eta_I$ intermediate states.  The detailed analysis has already been done by ref. ~\cite{Hambye:2009pw}
and they tell us the DM mass,  which satisfies the constraint of the correct relic density,
is about 534 GeV (in addition to the half of Higgs mass $\sim63$ GeV). 

{\it Direct detection}: Scattering between DM and nucleon can be used to detect DM and we have constraints on a spin-independent DM-nucleon scattering cross section  by several experiments such as XENON1T~\cite{XENON:2015gkh}.  
Since there is a mass difference between $\eta_R$ and $\eta_I$,
this is inelastic scattering. Thus,
we can simply suppress 
the t-channel scattering between DM and nucleon $N$ mediated by $Z$,  $\eta_R N \to \eta_I N$,  where the mass difference is taken to be more than $100$ keV to avoid this constraint.  
Since the mass difference arises from $\lambda''_{H\eta} v^2/(m_R+m_I)$ via Eq. (\ref{eq:mdif}), absolute value of the coupling $\lambda''_{H\eta}$ 
is required to be $5\times 10^{-7}\lesssim |\lambda''_{H\eta}|$ 
when the mass difference is larger than $100$ keV for the case that $m_R+m_I=300$ GeV.  Therefore,  this bound is very weak. 
The dominant contribution to DM-nucleon scattering is the one involving the Higgs boson mediation arising from interactions in the scalar potential.
However,  we simply assume the related coupling {$c_{\eta\eta h}$ for $\eta_R-\eta_R-h$ to be tiny 
enough; $c_{\eta\eta h} \lesssim 10^{-3}$}~\cite{Arcadi:2019lka},   in order 
to evade the direct detection via Higgs exchanges.

\section{Numerical analysis \label{sec:numerical}}
In this section,  we perform the numerical analysis.
Relevant free parameters $\{|y_{L'_{\mu}}|,  |y_{S_\mu}|,  |\lambda''_{H\eta}|,  m_{R},  M_{E, L'},  m_{E},   m_{I} \}$ are randomly scanned to search for the allowed region to satisfy muon $g-2$ anomaly
so that $\Delta a^{\rm new}_\mu$ matches with $\Delta a^{\rm exp}_\mu$. 
Simultaneously,  $\Delta S$ and $\Delta T$ in Eq.~(12) have to be within preferred oblique parameters in explaining  the W boson mass anomaly.  Then,  we check whether these allowed points can satisfy the constraint of $h\to \gamma\gamma$ decay width in Eq.~(19) or not. 
The ranges of the input free parameters are set to be as follows: 
\begin{align}
& [ |y_{L'_{\mu}}|,  |y_{S_\mu}|,  |\lambda''_{H\eta}|] \in [0. 1,  1],   \nn \\
&m_{R} \in [70, \ 550]\ [{\rm GeV}],  
\ \{M_{E, L'}\}\in [m_{R}, \ 5000]\ [{\rm GeV}],  \    \\
& m_{E} \in [10, \sqrt{4\pi}v]\ [{\rm GeV}],  \ m_I \in [m_{R}, \ \sqrt{m_R^2+ \sqrt{4\pi} v^2}]\ [{\rm GeV}] ,  \nn
\end{align}
where the lower bound on $m_R$ comes from LEP bound.  We take the upper bounds on $m_I$ and $m_E $ so that the perturbative limit for dimensionless couplings should be less than $\sqrt{4\pi}$, and
we choose $M_1 \le M_2$,  and take $m_{R} \le  (M_1,  m_I)$ so that $\eta_{R}$ is DM. 
Furthermore,   we work on the following regions $[|y_{L'_{e, \tau}}|,  |y_{S_{e, \tau}}|] \le 0. 001$ 
in order simply to evade the constraints of lepton flavor violations such as $\mu\to e\gamma$.  

\begin{figure}[tb]
\begin{center}
\includegraphics[width=97.0mm]{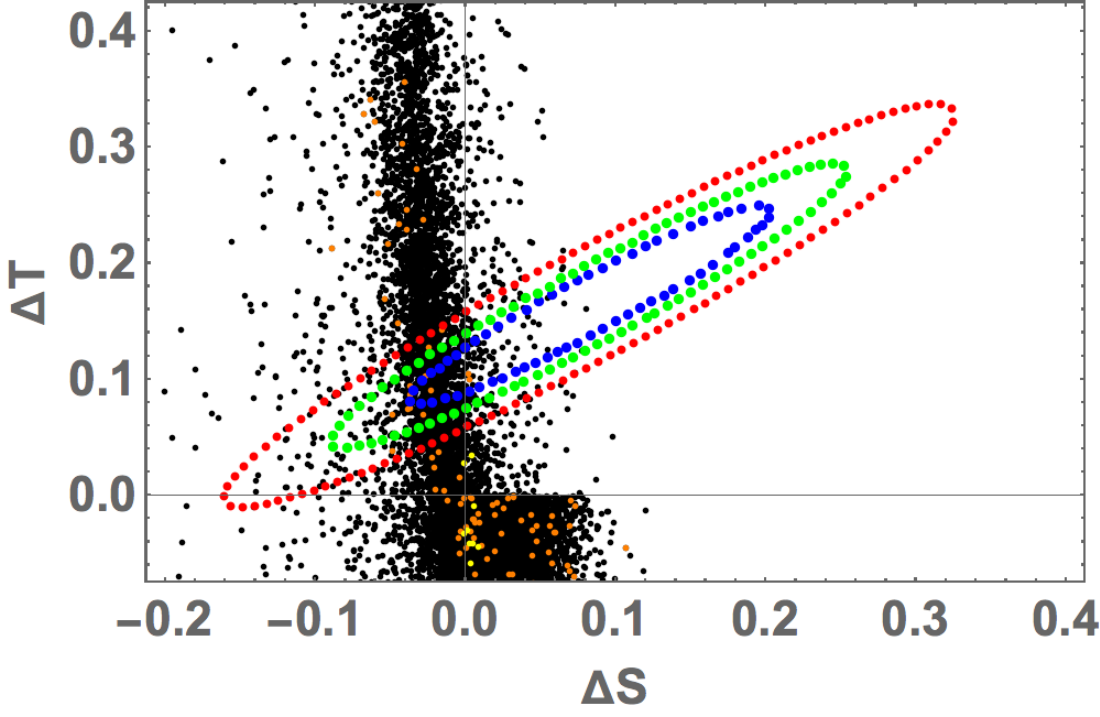} 
\caption{Oblique parameters $\Delta S$ and $\Delta T$ obtained in the model compared with the required region considering the W boson mass anomaly, 
where the blue,  green, and red circle represents the allowed region within $68\%$ confidence level (CL),   $90\%$ CL,  and $99\%$ CL.  The black dots are the ones satisfying the muon $g-2$ within 3$\sigma$ interval. 
The orange(yellow) dots satisfy the lighter(heavier) mass of DM at the region of $62. 5-63(525. 5-542. 5)$ GeV. } 
  \label{fig:st}
\end{center}\end{figure}

As a result of parameter scanning,  we obtain possible values of $\Delta S$ and $\Delta T$ for the parameter sets that can explain the deviation of muon $g-2$. 
Fig.~\ref{fig:st} shows oblique parameters $\Delta S$ and $\Delta T$ obtained in the model compared with the required region in light of the W boson anomaly, 
where the blue, green, and red circles respectively represent the allowed region within $68\%$ confidence level (CL), $90\%$ CL, and  $99\%$ CL. The black dots correspond to parameter sets satisfying the muon $g-2$ within 3$\sigma$ interval. 
We thus find that it is possible to obtain $\Delta S$ and $\Delta T$ which accommodate the W boson anomaly while explaining the deviation of muon $g-2$.
Here, we find more parameter sets giving $\Delta S < 0$. 
The orange(yellow) dots satisfy the lighter(heavier) mass of DM at the region of $62. 5-63(525. 5-542. 5)$ GeV in addition to the muon $g-2$ condition. 
From the figure, it is seen that the lighter DM would be in favor of the W boson mass anomaly. Hereafter,  we depict figures within the range of oblique parameters.

\begin{figure}[tb]
\begin{center}
\includegraphics[width=77.0mm]{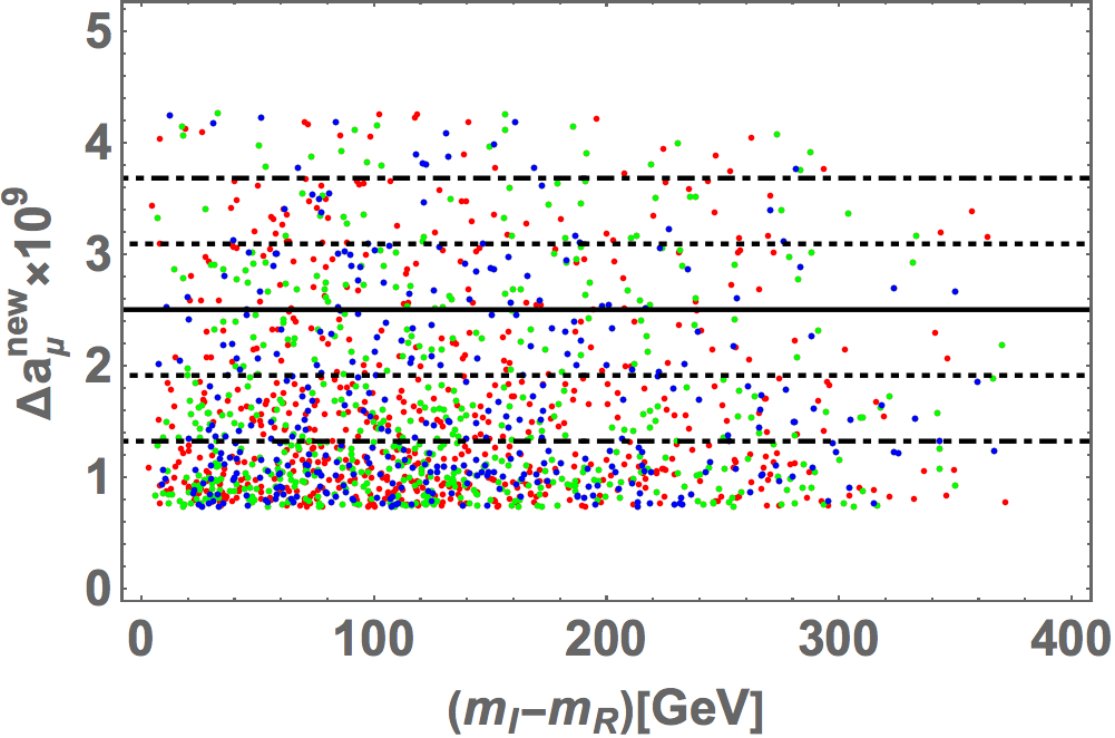} 
\includegraphics[width=77.0mm]{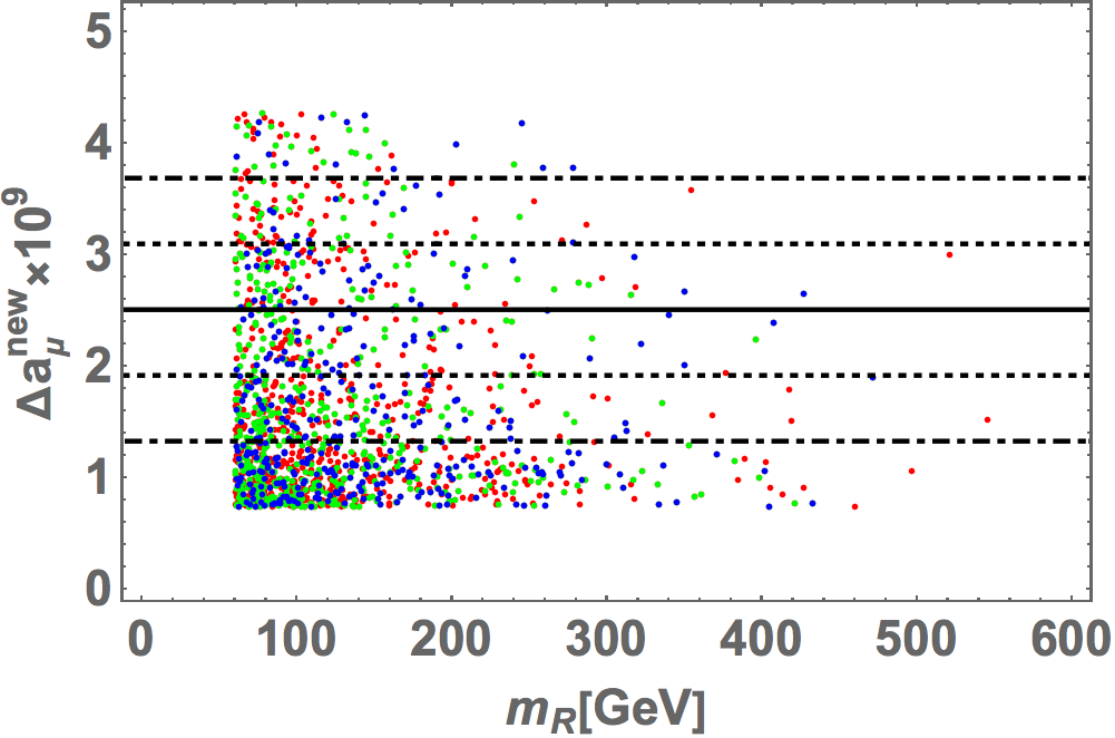} 
\caption{Figures show allowed mass range of neutral scalar bosons to obtain sizable muon $g-2$. 
The left(right) figure is muon $g-2$ at $3\sigma$ interval in terms of $m_I-m_R$($m_R$), 
where the red,  green, and blue dots correspond to the allowed region of oblique parameters within $68\%$ confidential level (CL),   $90\%$ CL,  and  $99\%$ CL.  
The solid black line represents the BF of muon $g-2$,  
the region between dotted lines is the $1\sigma$ one,  and the region between dot-dashed lines is the $2\sigma$ one. } 
  \label{fig:diff-damu}
\end{center}\end{figure}
%
We obtain the allowed range of new scalar particle masses to explain muon $g-2$ and W boson mass anomaly at the same time. 
The left(right) of Fig.~\ref{fig:diff-damu} shows muon $g-2$ at $3\sigma$ interval in terms of $m_I-m_R$($m_R$), 
where the red,  green, and blue dots correspond to the allowed region of oblique parameters within $68\%$ CL,   $90\%$ CL,  and $99\%$ CL.  
The solid  black line represents the best fit (BF) value of muon $g-2$,  
the region between dotted lines is $1\sigma$ {interval},  and the region between dot-dashed lines is the $2\sigma$ one. 
We then find the requirement for masses as $m_I - m_R \lesssim 370$ GeV and $m_R \lesssim 550$ GeV.  
The smaller mass difference tends to get smaller muon $g-2$,  which is a trivial result since the muon $g-2$ is proportional to the mass difference.  The smaller DM mass provides a wider allowed region that is also a natural consequence of the formula of muon $g-2$. 
When the DM mass and the mass difference are heavier,  the allowed space is narrower since they conflict with the oblique parameters.

\begin{figure}[tb]
\begin{center}
\includegraphics[width=77.0mm]{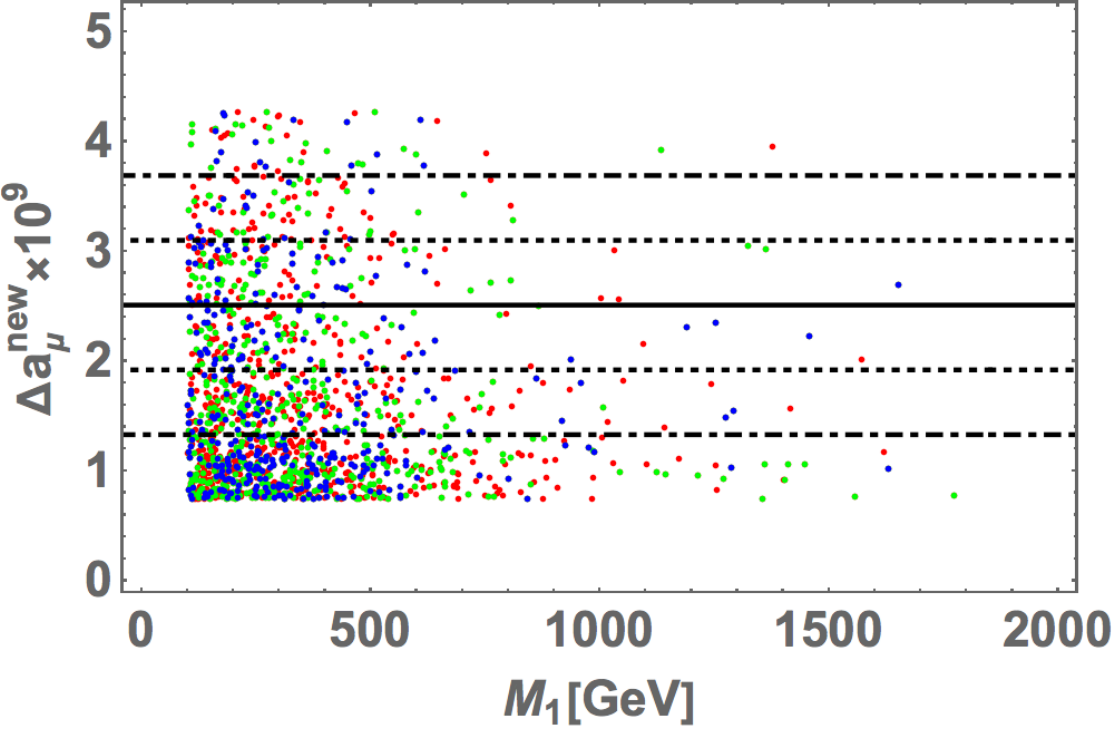} 
\includegraphics[width=77.0mm]{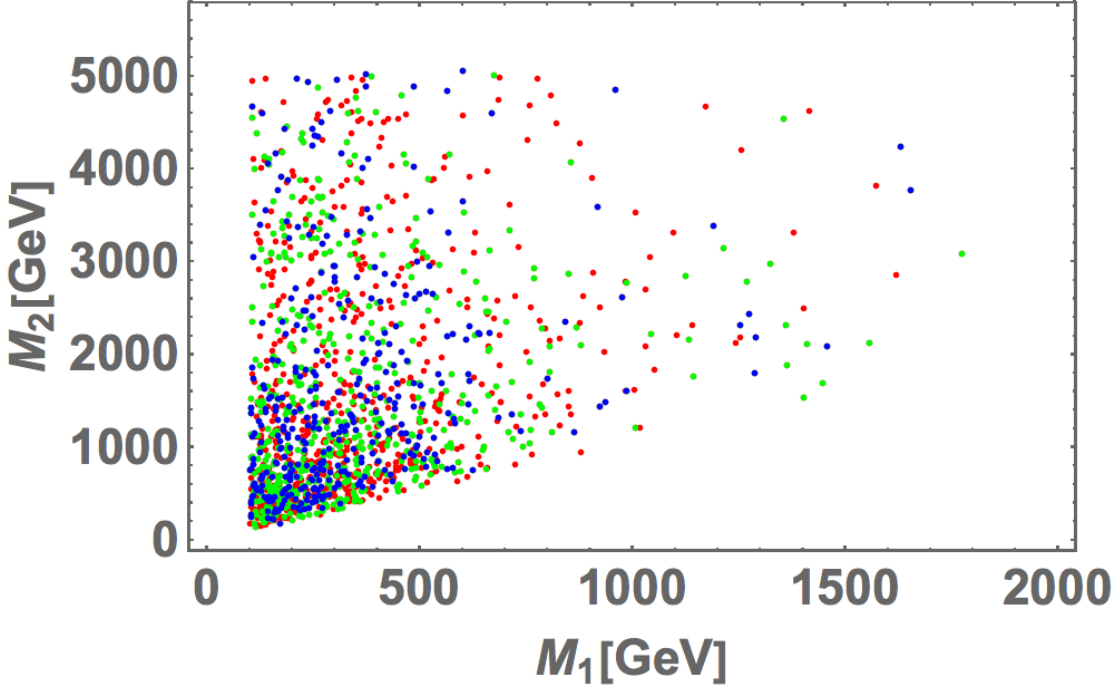} 
\caption{Figures show allowed mass range of new fermions to obtain sizable muon $g-2$. 
The left(right) figure is muon $g-2$($M_2$) at $3\sigma$ interval in terms of $M_1$, 
where the color and line legends are the same as in the Fig.~\ref{fig:diff-damu}. } 
  \label{fig:M12-damu}
\end{center}\end{figure}
%
We can also constrain extra fermion masses $M_1$ and $M_2$ when we require that the model explains muon $g-2$ and W boson mass anomaly. 
Left plot in Fig.~\ref{fig:M12-damu} shows muon $g-2$ at $3\sigma$ interval in terms of $M_1$ while the right plot shows a correlation between $M_1$ and $M_2$, 
where the color and line legends are the same as in the Fig.~\ref{fig:diff-damu}. 
{We then find $M_1 \lesssim 1800$ GeV while $M_2$ is not much constrained. 
Similar to {Fig.~\ref{fig:diff-damu}},  the smaller mass of $M_1$ tends to have more points, 
and the allowed mass ranges are found to be {$100 \ {\rm GeV}\lesssim M_1\lesssim 1800$ GeV and $100 \ {\rm GeV}\lesssim M_2\lesssim 5000$} GeV within the input ranges.  

\begin{figure}[tb]
\begin{center}
\includegraphics[width=77.0mm]{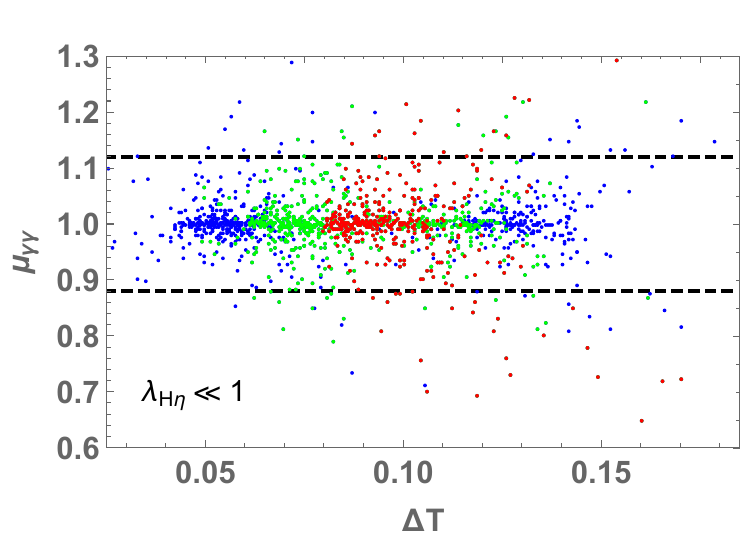} 
\caption{Predicted values of $\mu_{\gamma \gamma}$ in terms of $\Delta T$ with $\lambda_{H\eta} \ll 1$, 
where the indication of colors are the same as the Fig.~\ref{fig:diff-damu}. 
The region between the dashed line corresponds to the allowed region at 90$\%$ CL. 
} 
  \label{fig:diphoton}
\end{center}\end{figure}

In Fig.~\ref{fig:diphoton},  we show $\mu_{\gamma \gamma}$ for the allowed parameter sets as a function of $\Delta T$
 where the colors are the same as the Fig.~\ref{fig:diff-damu} and the region between the dashed line corresponds to a 90$\%$ CL allowed region {in Eq.~(\ref{eq:hto2photons})}. 
 Here we consider  $\lambda_{H\eta } \ll 1$ to see the effect of exotic charged leptons for illustration~\footnote{Note that we have a contribution from a one loop diagram in which the charged scalar boson propagates when we make the value of $\lambda_{H \eta}$ to be large.
 In this case, we would have cancellation between charged scalar and fermion loop diagrams, and 
 the points in the excluded region can be allowed by choosing the $\lambda_{H \eta}$ value.}.
The deviation from the SM ($\mu_{\gamma \gamma} =1$) can be sizable. 

Finally, in Tab.~\ref{bp1}, we show a
benchmark point
to satisfy the requirements from DM, muon $g-2$ within $3\sigma$ interval, oblique parameters within $68\%$ CL, 
and the W boson mass anomaly.
As expected from the sizable muon $g-2$,  rather large $|\lambda''_{H\eta}|(\sim -0. 865)$ is requested to satisfy the condition for oblique parameters at the same time.
Here we also comment on constraints from charged particle production at the collider experiments since $E_1$ is as light as $105$ GeV.  
At the LHC, it can be pair produced as $pp \to E_1^+ E_1^-$ via electroweak interactions.  Then $E_1$ decays into $\mu$ and $\eta_R$ via Yukawa interaction.  As a constraint,  we refer to a search for slepton pair production in which slepton decays into charged lepton and neutralino (DM candidate in a supersymmetric model)~\cite{ATLAS:2019lff}.  
We find that $105$ GeV slepton and $\sim 60$ GeV neutralino region is still allowed,  and the benchmark point is still safe from collider constraint. 
Heavier charged particle has more complicated decay chains 
and we need detailed analysis to obtain discovery potential at collider experiments. 
A more detailed discussion of collider physics is beyond the scope of this work and it is left as future work.

\begin{table}[th]
	\centering
	\begin{tabular}{|c|c|c|} \hline 
			\rule[14pt]{0pt}{0pt}
		$[m_R, m_I,  m_{\eta^\pm}]/{\rm GeV} $ & $[62. 9,  237,  288]$   \\ \hline
		\rule[14pt]{0pt}{0pt}
		$[M_E, M_{L'},  m_E]/{\rm GeV}$ & $[1101,  124. 4,  137. 7]$     \\ \hline
		\rule[14pt]{0pt}{0pt}
%
		$[M_1, M_2]/{\rm GeV}$ &  $[105. 4,  1120]$    \\ \hline
		\rule[14pt]{0pt}{0pt}
		$[y_{L'_\mu},  y_{S_\mu}]$ & $[0. 0194,  -0. 307]$     \\ \hline
		\rule[14pt]{0pt}{0pt}
		$\lambda''_{H\eta}$  &  $-0. 865$ \\ \hline
		\rule[14pt]{0pt}{0pt}
		$\langle\sigma v_{\rm rel}\rangle_Y\ {\rm /GeV^{-2}}$ &  $1. 29\times10^{-14}$        \\ \hline
		\rule[14pt]{0pt}{0pt}
		$ \Delta a^{new}_{\mu}$ & $ 9. 73\times 10^{-10}$   \\ \hline
		\rule[14pt]{0pt}{0pt}
		$\Delta T$ &  $ 0. 100$   \\ \hline
		\rule[14pt]{0pt}{0pt}
		$\Delta S$ &  $ 0. 00248$   \\ \hline
		 $\mu_{\gamma \gamma}$  &   $ 0. 93$  \\ \hline
				\hline
	\end{tabular}
	\caption{Numerical benchmark points  to satisfy DM,  muon $g-2$ at $3\sigma$ interval,  and oblique parameters at $68\%$ CL. }
	\label{bp1}
\end{table}
%

\section{Summary and Conclusions}
 We have successfully explained the W boson mass anomaly reported by CDFII collaboration and muon $g-2$ reported by BNL both of which suggests NP, at the same time. 
For this purpose,  we have introduced exotic fields; one isospin doublet vector fermion,  one isospin singlet singly-charged vector fermion,  and an isospin doublet inert scalar field {imposing $Z_2$ odd parities for these new fields}.  
These new particles also provide a bosonic DM candidate as well as sizable muon $g-2$ with no chiral suppression at one-loop level.

Our findings are as follows.  In order to obtain the sizable muon $g-2$,  we need nonzero mass squared difference between $\eta_R$ and $\eta_I$ that is clearly found in Eq.~(15). 
The exotic vector-like fermions negatively contribute to $\Delta T$ (as well as $\Delta S$) while the doublet scalar can provide a positive contribution.  
Because the W boson anomaly can be explained by a positive shift of $\Delta T$ as shown in Fig.~\ref{fig:st},  the doublet scalar contribution is more important. 
%
Taking into consideration of the above features,  we have performed the numerical analysis,  and found an allowed parameter region,  as shown in Figs.~\ref{fig:diff-damu} and \ref{fig:M12-damu}. 
In addition,  we have found that $\eta_R$ can be a DM candidate that satisfies the observed relic density and constraints of direct searches. 
Taking into consideration of DM, it implies that muon $g-2$ tends to be a bit lower than the BF value. But it is still within 3$\sigma$ interval.
\section*{Acknowledgments}
KIN was supported by JSPS Grant-in-Aid for Scientific Research (A) 18H03699,  (C) 21K03562,  (C) 21K03583,  Okayama Foundation for Science and Technology,  and Wesco Scientific Promotion Foundation. 
The work of H. O.  is supported by an appointment to the JRG Program at the APCTP through 
the Science and Technology Promotion Fund and Lottery Fund of the Korean Government,  
and also by the Korean Local Governments - Gyeongsangbuk-do Province and Pohang City 
H.  O.  is sincerely grateful for the KIAS members.  
The work was supported in part by the Fundamental Research Funds for the Central Universities (T. ~N. ). 

\begin{appendix}

\section{Vector-like fermion contributions to electroweak vacuum polarization diagrams}
\label{appdx:polarizationdiagrams}
To estimate the oblique parameters,  we evaluate the electroweak vacuum polarization diagrams with vector-like fermions 
and then summarize the analytic form of the contributions.  
The contributions of vacuum polarizations for $Z$ and $W$ to the oblique parameters are given 
\begin{align}
& \Pi_Z^{\mu \nu} = g^{\mu \nu} \frac{e^2}{c_W^2 s_W^2} (\Pi_{33}(q^2) - 2 s_W^2 \Pi_{3Q} (q^2) 
- s_W^4 \Pi_{QQ}(q^2)),  \\
& \Pi^{\mu \nu}_W = g^{\mu \nu} \frac{e^2}{s_W^2} \Pi_\pm(q^2), 
\end{align}
where $q$ is four momentum carried by gauge bosons. 
Those with the non-zero contribution to $\Pi_{33} (q^2)$ are listed below:
\begin{align}
& \Pi^{E_1E_1}_{33}(q^2) = - \frac{s_c^4}{16 \pi^2} F(q^2,  M_1^2, M_1^2),  \\
& \Pi^{E_2E_2}_{33}(q^2) = - \frac{c_c^4}{16 \pi^2} F(q^2,  M_2^2, M_2^2),  \\
& \Pi^{E_1E_2}_{33}(q^2) = - \frac{s_c^2 c_c^2}{8 \pi^2} F(q^2,  M_1^2, M_2^2),  \\
& \Pi^{N'N'}_{33}(q^2) = - \frac{1}{16 \pi^2} F(q^2,  M_{N'}^2, M_{N'}^2),  
\end{align}
where the superscripts on the left side represent the particles inside the vacuum polarization diagrams. 
$F(q^2, m^2, m'^2)$ is the loop function given by
\begin{align}
F(q^2, m^2, m'^2) = &  \int_0^1 dx dy \delta(1-x-y) \left(\frac{1}{\epsilon_{\rm \overline{MS}}} - \ln \left(\frac{\Delta}{\mu^2} \right) \right) \nonumber \\
& \times (2x (1-x) q^2 - x m^2 - y m'^2 + m m'),  \\
\Delta = & -x (1-x) q^2 + x m^2 + y m'^2,  \quad  \frac{1}{\epsilon_{\rm \overline{MS}}} \equiv \frac{2}{\epsilon} - \gamma - \ln(4 \pi), 
\end{align} 
where $\mu$ is auxiliary parameter with
mass dimension.  
Once the $S, T$-parameters are computed
from Eqs.  (\ref{eq:S}) and (\ref{eq:T}) in Sec.~\ref{sec:oblique},  the $\mu$ dependence cancels out and thus does not remain. 

In a similar way,  one finds a non-zero contribution for $\Pi_{3Q,  QQ,  \pm}$:
\begin{align}
& \Pi^{E_1E_1}_{3Q}(q^2) = - \frac{s_c^2}{8 \pi^2} F(q^2,  M_1^2, M_1^2),  \\
& \Pi^{E_2E_2}_{3Q}(q^2) = - \frac{c_c^2}{8 \pi^2} F(q^2,  M_2^2, M_2^2),  
\end{align}
\begin{align}
& \Pi^{E_1E_1}_{QQ}(q^2) = - \frac{1}{4 \pi^2} F(q^2,  M_1^2, M_1^2),  \\
& \Pi^{E_2E_2}_{QQ}(q^2) = - \frac{1}{4 \pi^2} F(q^2,  M_2^2, M_2^2),  
\end{align}
\begin{align}
& \Pi^{E_1N'}_{\pm}(q^2) = - \frac{s_c^2}{8 \pi^2} F(q^2,  M_1^2, M_{N'}^2),  \\
& \Pi^{E_2N'}_{\pm}(q^2) = - \frac{c_c^2}{8 \pi^2} F(q^2,  M_2^2, M_{N'}^2). 
\end{align}

Then the $S$- and $T$-parameters can be obtained from 
Eqs.  (\ref{eq:S}) and (\ref{eq:T}) in Sec. ~\ref{sec:oblique} and
those equations show that the divergent part proportional to $\epsilon_{\rm SM}$ is canceled if the oblique parameters are computed. 
\end{appendix}

\section*{Data Availability Statement}
The data that support the findings of this study are available on request from the authors.

\end{document}